\documentclass{article}
\usepackage{spconf,amsmath,graphicx}
\usepackage{amssymb}
\usepackage{multirow} 
\usepackage{CJKutf8} 
\usepackage{url} 

\usepackage{caption}
\usepackage{subcaption}
\usepackage{flushend}

\usepackage{subcaption}
\usepackage{booktabs}

\title{frame-level multi-label playing technique detection using multi-scale network and self-attention mechanism}
\name{Dichucheng Li $^1$, Mingjin Che $^2$, Wenwu Meng $^2$, Yulun Wu $^1$, Yi Yu $^3$, Fan Xia $^2$ and Wei Li $^{1,4}$\thanks{This work was jointly supported by NSFC(62171138), Key Laboratory of Intelligent Processing Technology for Digital Music (Zhejiang Conservatory of Music), Ministry of Culture and Tourism (Grant Number 2022DMKLB002). Wei Li, Fan Xia and Yi Yu are corresponding authors.}}
\address{$^1$ School of Computer Science and Technology, Fudan University, Shanghai, China \\
	$^2$ College of Experimental Art, Sichuan Conservatory of Music, Sichuan, China\\
    $^3$ Digital Content and Media Sciences Research Division, National Institute of Informatics (NII), Japan\\
    $^4$ Shanghai Key Laboratory of Intelligent Information Processing, Fudan University, China\\
 }

\begin{document}
\begin{CJK*}{UTF8}{gbsn}
	\ninept
	\maketitle
	\begin{abstract}
	Instrument playing technique (IPT) is a key element of musical presentation. However, most of the existing works for IPT detection only concern monophonic music signals, yet little has been done to detect IPTs in polyphonic instrumental solo pieces with overlapping IPTs or mixed IPTs. In this paper, we formulate it as a frame-level multi-label classification problem and apply it to Guzheng, a Chinese plucked string instrument. We create a new dataset, Guzheng\_Tech99, containing Guzheng recordings and onset, offset, pitch, IPT annotations of each note. Because different IPTs vary a lot in their lengths, we propose a new method to solve this problem using multi-scale network and self-attention. The multi-scale network extracts features from different scales, and the self-attention mechanism applied to the feature maps at the coarsest scale further enhances the long-range feature extraction. Our approach outperforms existing works by a large margin, indicating its effectiveness in IPT detection.
	
	\end{abstract}

	\section{Introduction}
	\label{sec:intro}
\begin{figure}[t]
    \centering
    \includegraphics[width=8.5cm]{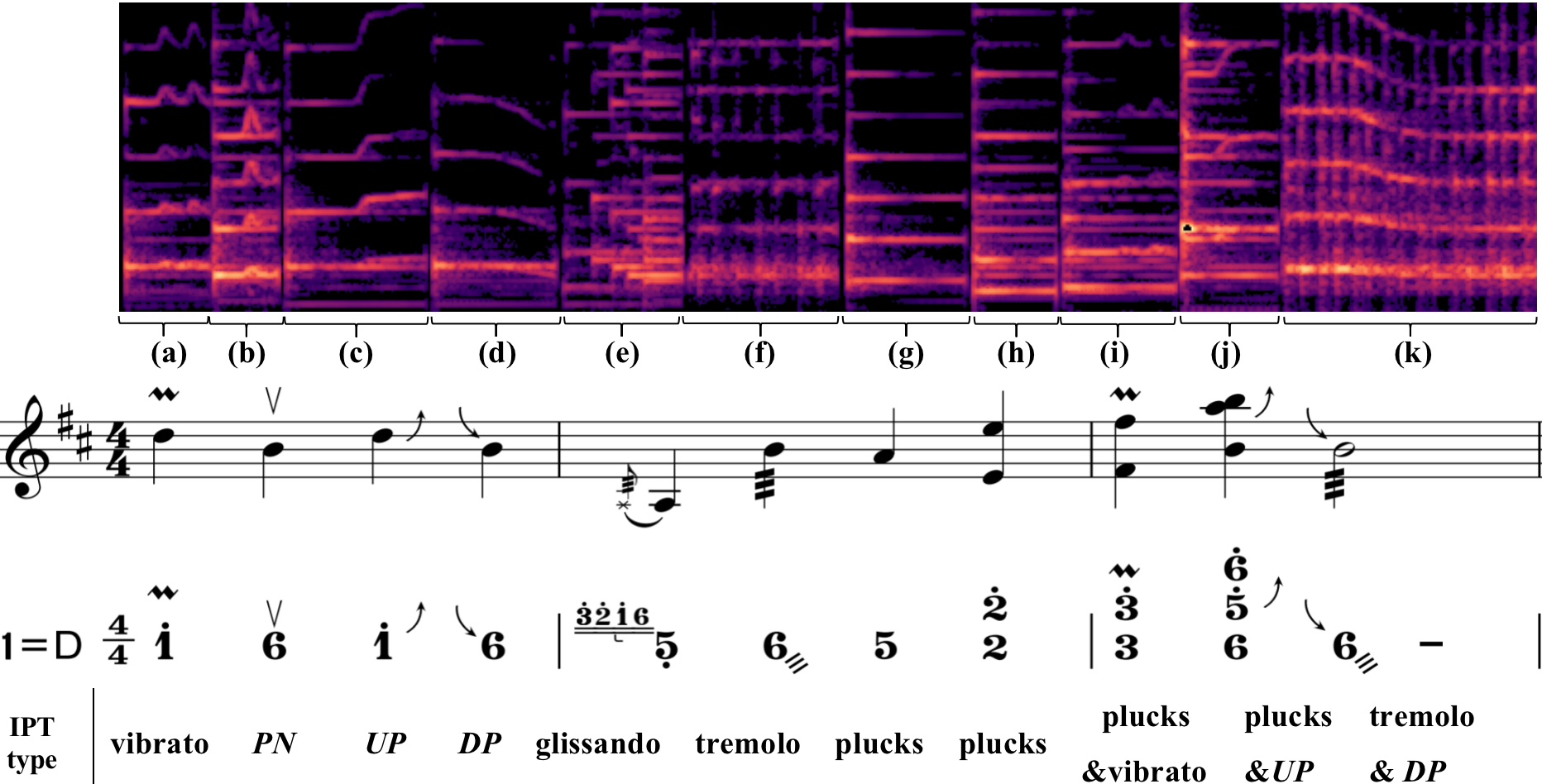}
    \caption{The spectrogram, staff notation, numbered musical notation and IPT annotation of a Guzheng phrase. ``IPT", ``{\itshape UP}", ``{\itshape DP}", ``{\itshape PN}" denote ``Instrument playing technique",  ``Upward Portamento", ``Downward Portamento" and ``Point Note" respectively.}
    \label{fig:notation}
    \end{figure}

    Instrument playing technique (IPT) is a key element in enhancing the vividness of musical performance. As shown by the Guzheng numbered musical notation (a musical notation system widely used in China) in Fig.\ref{fig:notation}, a complete automatic music transcription (AMT) system should contain IPT information in addition to pitch and onset information. IPT detection aims to classify the types of IPTs and locate the associated IPT boundaries in audio. IPT detection and modeling can be utilized in many applications of music information retrieval (MIR), like performance analysis \cite{yang2017computational} and AMT \cite{TENT}. 
    
	The research on IPT detection is still in its early stage. Existing works mostly used machine learning methods, combined with hand-crafted features \cite{wang2020playing, wang2022adaptive}. Early studies in this field mainly focused on the IPT classification of isolated notes \cite{abe2010feature, su2014sparse}. However, there are always multiple notes with varying IPTs in an audio sequence and we not only need to classify the IPTs but also locate their boundaries. In \cite{TENT, chen2015electric}, a two-step method was proposed to detect IPTs in 42 electric guitar solo tracks. Firstly, the positions of IPT candidates were located in the extracted melody contours. Then, hand-crafted features of the candidates were fed to classifiers such as support vector machine. However, the robustness of the method may be degraded by errors caused in melody extraction, the intermediate step. 

    With the advancements in deep learning, deep neural networks have been increasingly used in more recent work \cite{guqin, liang2019piano}. In \cite{foldedcqt}, a convolutional recurrent neural network (CRNN) based model was proposed to classify IPTs in audio sequences concatenated by cello notes from 5 sound banks. To alleviate the computational redundancy caused by the sliding window in \cite{foldedcqt}, Wang et al. \cite{wang2019musical} proposed a fully convolutional network (FCN) based end-to-end method to detect IPTs in segments concatenated by isolated Erhu notes. In \cite{li2022playing}, an additional onset detector was used, and its output was fused with IPT prediction in a post-processing step to improve the accuracy of IPT detection from monophonic audio sequences concatenated by isolated Guzheng notes. However, Guzheng is a polyphonic instrument. In Guzheng performance, notes with different IPTs are usually overlapped (region (i), (j) of Fig.\ref{fig:notation}) and mixed IPTs that can be decomposed into multiple independent IPTs are usually used (region (k) of Fig.\ref{fig:notation}). Thus, the frame-level single-label IPT detection model can hardly be applied directly to music with overlapping IPTs or mixed IPTs, like harp or Guzheng solo pieces.
    
    Annotating IPTs in music containing overlapping or mixed IPTs is labor-intensive, so most existing work on IPT detection typically uses datasets with monophonic instrumental solo pieces \cite{TENT, wang2020playing, guqin} or audio sequences concatenated by individual IPT clips \cite{foldedcqt, wang2019musical, li2022playing}. However, these datasets lack the complexity present in music with overlapping IPTs or mixed IPTs. The randomly generated sequences are even not natural enough, as they ignore music theory, rests, and variations in timbre and intensity in real recordings. As shown in \cite{wang2019musical, li2020instrument}, models that perform well on such sequences may have insufficient generalization capabilities when applied to real recordings. 

    To overcome the limitations mentioned above, we formulate the problem as a frame-level multi-label classification problem. Our study focuses on Guzheng, a plucked 21-string Chinese instrument with diverse playing techniques, but our proposed methods can be applied to other instruments as well. We create a new dataset, Guzheng\_Tech99, containing 99 Guzheng solo pieces of various genres recorded by professional Guzheng players, along with 63,352 annotated labels indicating the onset, offset, pitch and playing techniques of every note in each recording. The lengths of different IPTs vary a lot. Long-range features are crucial for long IPTs, while high-resolution features are necessary for short IPTs due to subtle variation. From this characteristic of IPTs, we propose a method using multi-scale convolution and self-attention. The multi-scale network extracts features from different scales, and self-attention blocks at the coarsest scale enhance the extraction of long-range features. This is the first approach, to our knowledge, to detect IPTs in instrumental solo pieces with overlapping or mixed IPTs, a more complex and general scenario for IPT detection. More details about the code and datasets can be found at \url{https://lidcc.github.io/GuzhengTech99/}.
	
	\section{DATASET}
    In this section, we introduce the Guzheng playing techniques considered in our work and the process of making the dataset.
    \subsection{Playing Technique Descriptions} 
    Traditional Guzheng playing techniques can be categorized into two classes: plucking strings with the right hand and pressing strings with the left hand \cite{IPTclass}. In practice, left-hand and right-hand IPTs are often used simultaneously. As shown in Fig.\ref{fig:notation}, this can occur in two cases: when notes with different IPTs are overlapped (region(i), (j) of Fig.\ref{fig:notation}) or when a note is played with a mixed IPT (region(k) of Fig.\ref{fig:notation}) that can be decomposed into multiple independent IPTs based on the Guzheng numbered musical notation. Additionally, we consider several notes played simultaneously with the same IPT as a single IPT (region(h) of Fig.\ref{fig:notation}). Consequently, we consider seven independent IPTs (see Table \ref{tab:statistic}) in our work.
    
    We categorize left-hand playing techniques into four types that produce different pitch variations. \textbf{Vibrato} (chanyin颤音) produces periodic pitch changes based on a base pitch. The pitch of \textbf{Point Note} (dianyin点音, {\itshape PN} for short hereinafter) is raised briefly right after the note beginning and then restored instantly. \textbf{Upward Portamento} (shanghuayin上滑音, {\itshape UP} for short hereinafter) slides the pitch continuously from low to high, while \textbf{Downward Portamento} (xiahuayin下滑音, {\itshape DP} for short hereinafter) slides the pitch continuously from high to low. As regards to the three right-hand playing techniques, we define \textbf{Plucks} (gou勾, da打, mo抹, tuo托, cuo撮, etc.) as normally plucking one or more strings without special playing techniques. \textbf{Glissando} (guazou刮奏, huazhi花指, etc.) is a rapid slide across discrete pitches in pentatonic scales. \textbf{Tremolo} (yaozhi摇指) is a rapid reiteration of a single note.

    \begin{table}
 \begin{center}
 \begin{tabular}{|c|c|c|c|c|c|}
  \hline
  \multirow{2}{*}{IPT} & \multirow{2}{*}{num} & \multicolumn{4}{c|}{length (seconds)} \\
  \cline{3-6}
    &   & sum & mean & max & min\\
  \hline
  vibrato & 1994 & 1650.31 & 0.83 & 4.37 & 0.21\\
  \hline
  {\itshape UP} & 756 & 544.12 & 0.72 & 3.84 & 0.10\\
  \hline
  {\itshape DP} & 208 & 126.56 & 0.61 & 3.44 & 0.19\\
  \hline
  {\itshape PN} & 209 & 153.12 & 0.73 & 3.24 & 0.23\\
  \hline
  glissando & 734 & 67.54 & 0.09 & 0.39 & 0.03\\
  \hline
  tremolo & 77 & 152.75 & 1.98 & 4.67 & 0.21\\
  \hline
  plucks & 11860 & 7066.19 & 0.60 & 6.82 & 0.07\\
  \hline
 \end{tabular}
\end{center}
 \caption{Detailed statistics of each IPT in the dataset. The columns of ``num", ``sum", ``mean", ``max", ``min" indicate the note number, the total length, average length, maximum length, minimum length.}
 \label{tab:statistic}
\end{table}

    \subsection{Data Collection and Labelling}
    The proposed dataset, Guzheng\_Tech99, consists of 99 audio recordings of Guzheng solo compositions recorded by two professional Guzheng players in a professional recording studio. The Guzheng for recording is a modern, full-sized Guzheng with 21 nylon-coated steel strings and the bass strings wound in copper. The audio excerpts in the dataset cover most of the genres of Guzheng music that vary in composing and performing styles, including Henan, Shandong, Chaozhou, Zhejiang, Shaanxi, and so on. The audio recordings in the dataset are 9064.6 seconds long in total. 

    We label the onset, offset, pitch and playing techniques of every note in each recording by {\itshape Sonic Visualiser} \cite{sonic_visualiser}. We treat the series of discrete tones in glissando as several independent notes while treating the rapid reiteration of a single tone of tremolo as a whole note according to the Guzheng numbered musical notation. As a result, the dataset consists of 63,352 annotated labels in total. Table \ref{tab:statistic} shows the detailed statistics of each IPT in the dataset. The time scope that is not covered by our annotation is the rests in music or the blank parts at the beginning and end of the recording. 
    Finally, the track-level information (audio id, audio name, mode, time signature, performer, genre, and audio length) is collected in a metadata file. 

    The dataset is split into 79, 10, 10 songs respectively for the training set, the validation set, and the test set (roughly 7330 seconds of audio for training, 819 seconds for validation, and 916 seconds for testing). When splitting, we control the distribution of IPT types and performers in the three sets to be as similar as possible. 

    \begin{figure*}[t]
    \centering
    \includegraphics[width=17.5cm]{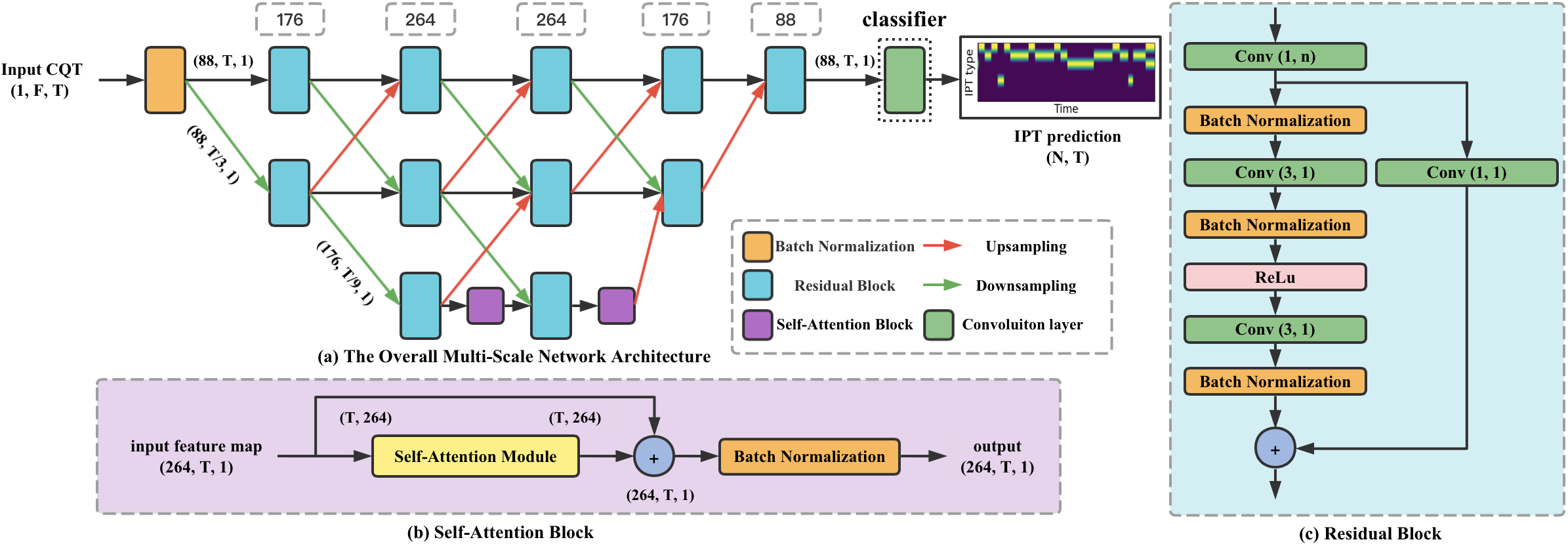}
    \caption{The overall multi-scale network architecture of our proposed model (a), the detailed structure of the self-attention block (b) and the residual block (c). The numbers in dashed boxes on the top row of (a) is the output channel number of the residual blocks in that column. The numbers in parentheses in (a), (b) indicate the shape of the feature maps at the corresponding positions, and the numbers in parentheses in (c) denote the kernel size of the convolution layer, in which ``n" equals to the number of the last dimension of the residual block input. ``F", ``T" denote the length of the frequency axis and time axis of the input. ``N" is the number of the IPT types.}
    \label{fig:model}
    \end{figure*}
    \section{Method}
    The overall framework of our proposed model is illustrated in Fig.\ref{fig:model}. The multi-scale network forms the overall architecture. The residual block and the self-attention block are the main modules.
    \subsection{Input Representation}
    Constant-Q Transform (CQT) is used as the input feature representation of audio in our experiments. The frequency scale of CQT is musically and perceptually motivated \cite{cqt}. Raw audio was clipped to 3-second pieces and resampled to 44.1 kHz. We use {\itshape librosa} \cite{librosa} to compute CQT with sampling rate of 44,100, hop length of 512, bins per octave of 12, fmin of 27.5 Hz, frequency bins number of 88.
    
    \subsection{Multi-Scale Network}
 
    Inspired by the success of convolutional neural networks (CNN) in many MIR tasks \cite{liang2019piano, hung2018frame}, we apply CNN in our task. However, the kernels of CNNs are designed to capture local information, thus a single convolution layer has a limited receptive field. As shown in Table \ref{tab:statistic}, different Guzheng playing techniques vary a lot in their lengths (the mean length of glissando notes is 0.09 seconds, while that of tremolo notes is 1.98 seconds) and there are also large differences in length within the same kind of playing techniques (the maximum and minimum length of plucks is 6.82 and 0.07 seconds). If only a certain part of an IPT is considered, it tends to misjudge one IPT as another. For example, a portamento note will be misjudged as a normal pluck note if its inflection point in the spectrogram is out of the receptive field of the model. As solutions, the receptive field is usually enlarged by directly stacking multiple convolution layers or use large-size convolution kernels. But these methods both result in an excess of parameters. Simply increasing the receptive field will also lead to a loss in details for the subtle change of short IPTs. 
    
    To address the issue, we introduce the multi-scale network, which was firstly proposed in computer vision tasks \cite{wang2020deep}. As shown in Fig.\ref{fig:model}(a), our proposed model is composed of three horizontal branches for different scales in the time axis. The resolution of the feature in the branches from top to bottom is from high to low. The middle branch with the medium resolution is used as a transition for the fusing between high-resolution features and long-range features. By downsampling/upsampling the feature to different scales, long-range features can be fused with high-resolution features repeatedly.

    To convert the spectral information into the channel domain, we first process the CQT input (1, F, T) into a sequence with the shape of (88, T, 1) by reshaping and batch normalization. Through the whole architecture, multi-scale fusion was repeated by rescaling and concatenation. The rescaling procedure is achieved by max-pooling layers and transposed convolution layers with kernel size of 3 × 1. After rescaling the feature maps from different resolutions to a unified scale, they are concatenated along the last dimension. 

    Fig.\ref{fig:model}(c) is the structure of the residual block. It is based on \cite{hung2018frame, chou2018learning}, but a convolution layer is added as the first layer to adjust the last dimension of the input to 1. The skip connection in the residual block is designed to help the model learn deeper. 
    
    After the feature processing, we obtain a feature map with the shape of (88, T, 1). Then, a 3 × 1 convolution layer followed by a sigmoid layer outputs a prediction with the shape of (7, T) for representing the likelihood of the presence of each IPT per frame.
    To decide the existence of an IPT, we pick a threshold of 0.5 to binarize the result. We use the weighted binary cross entropy (BCE) proposed in \cite{citeloss} as the loss function to counter class imbalance. We fine-tune the model using stochastic gradient descend (SGD) with momentum 0.9, an initial learning rate of 0.01, a batch size of 10, a gradient clipping L2-norm of 3, and a cosine learning rate scheduler.
    
	\subsection{Self-Attention Block}
    Since temporal modeling is crucial in IPT detection task, we apply self-attention mechanism, which has been proven effective in many sequence modeling tasks \cite{vaswani2017attention, wu2020multi}. As shown in Fig.\ref{fig:model}, the self-attention blocks applied to the feature maps at the coarsest scale further enhance the extraction of global features by capturing interactions between different frames on the feature maps. For an input audio sequence $X = (x_1, x_2, …, x_T)$ of length $T$, with $x_t \in \mathbb{R}^{d_m} $. The input $X$ is transformed into namely ``queries" $Q$ $\in \mathbb{R}^{T × d_k}$, ``keys" $K$ $\in \mathbb{R}^{T × d_k}$ and ``values" $V$ $\in \mathbb{R}^{T × d_k}$. Each element of them is calculated as Eq.1. 
    
    \begin{equation}
    \begin{split}
    q_i = x_iW_Q, \\
    k_i = x_iW_K, \\
    v_i = x_iW_V.
    \end{split}
    \end{equation}$W_Q$, $W_k$, $W_v$ $\in \mathbb{R}^{d_m × d_k}$ are three trainable parameters to calculate the ``queries", ``keys" and ``values", respectively. As shown in Eq.2, each element of the output sequence $O = (o_1, o_2, …, o_T)$, where $o_i \in \mathbb{R}^{d_k}$, is computed as a weighted sum of the ``values” elements, and the results are fed to a softmax computation. The weight of each ``values" element is computed by a scaled dot-product between the corresponding ``queries" element and ``keys" element.
    
    \begin{equation}\label{relativity}
    o_i = softmax(\sum_{j=1}^{T}\frac{q_ik_j^\mathrm{T}}{\sqrt{d_k}}v_j).
    \end{equation}

    The output of the self-attention module was reshaped, added with the input and then passed through a batch normalization layer to get the final output of the self-attention block.

	\section{Experiments}
    We use precision $P = \frac{TP}{TP+FP}$, recall $R = \frac{TP}{TP+FN}$, and F1-score $F1 = \frac{2PR}{P+R}$ as the evaluation metrics, where $TP$, $FP$, $FN$ denote true positives, false positives, and false negatives, respectively. The output of the model is compared to the ground truth labels per frame.

    \begin{table}
        \begin{center}
        \renewcommand\arraystretch{1}
        \setlength\tabcolsep{5mm}{
         \begin{tabular}{c|ccc}
          \hline
          Method &  precision & recall & F1-score\\
          \hline\hline
          w/o self-att & 85.43 & 82.64 & 84.01 \\
          w/o res & 86.38 & 83.45 & 84.89 \\
          Single-scale & 79.04 & 75.02 & 76.98 \\
          \hline
          Proposed & \textbf{87.62} & \textbf{85.48} & \textbf{86.54} \\
          \hline
        \end{tabular}
        }
     \end{center}
     \caption{Ablation studies with frame-level precision, recall and F1-score on Guzheng\_Tech99 dataset.}
     \label{tab:ablation}
    \end{table}

    \begin{table}
        \begin{center}
        \renewcommand\arraystretch{1}
        \setlength\tabcolsep{4.5mm}{
         \begin{tabular}{c|ccc}
          \hline
          Method &  precision & recall & F1-score\\
          \hline\hline
          \texttt{GZFNO} \cite{li2022playing} & 69.49 & 68.00 & 68.73 \\
          \texttt{EHFCN} \cite{wang2019musical} & 69.55 & 67.77 & 68.65 \\
          \texttt{CNN+Res} \cite{chou2018learning} & 75.97 & 66.23 & 70.77 \\
          \hline
          Proposed & \textbf{87.62} & \textbf{85.48} & \textbf{86.54} \\
          \hline
        \end{tabular}
        }
     \end{center}
     \caption{Results of the proposed and baseline methods with frame-level precision, recall and F1-score on Guzheng\_Tech99 dataset.}
     \label{tab:compare}
    \end{table}

	\subsection{Ablation Study}
	We perform an ablation study to evaluate the effectiveness of each part in our proposed model.
    Firstly, we remove the self-attention blocks in the model. As shown in the first row (\texttt{w/o self-att}) of Table \ref{tab:ablation}, the precision, recall, and F1-score decreased by 2.19\%, 2.84\%, and 2.53\%, respectively, demonstrating that the self-attention mechanism effectively captures long-range dependencies to improve the performance of the model. Then, we remove the skip-connection in the residual blocks. As shown in the second row (\texttt{w/o res}) of Table \ref{tab:ablation}, the precision, recall, and F1-score were reduced by 1.24\%, 2.03\%, and 1.65\%, respectively, proving the effectiveness of the residual blocks in our model. Finally, we alter the architecture to a single-scale one by removing the whole downsampled subnetworks and only retaining the one with the highest resolution (the topmost branch in Fig.\ref{fig:model}(a)). As shown in the third row (\texttt{Single-scale}) of Table \ref{tab:ablation}, the precision, recall and F1-score all drop drastically: 8.58\%, 10.46\% and 9.56\% respectively, indicating the multi-scale network can better fuse long-range features with high-resolution features and improves the classification ability of our model.
 
    \subsection{Comparing with Existing Methods}
    Our proposed method is the first (to the best of our knowledge) to achieve IPT detection from instrumental solo pieces with overlapping IPTs or mixed IPTs, which is considered as a frame-level multi-label classification problem. To evaluate its performance in this task, we choose three models that achieved high performance in similar tasks as baseline methods: \texttt{GZFNO} \cite{li2022playing}, \texttt{EHFCN} \cite{wang2019musical} and \texttt{CNN+Res} \cite{chou2018learning}. Among them, \texttt{GZFNO} and \texttt{EHFCN}  are models that achieved great performance in frame-level single-label IPT detection task. \texttt{CNN+Res} shows great performance in sound event detection \cite{chou2018learning} and instrument recognition \cite{hung2018frame}, both of which are frame-level multi-label detection tasks. For \texttt{GZFNO} and \texttt{EHFCN}, we change the activation function of the last layer from softmax to sigmoid because softmax is used in multi-class single-label classification tasks while sigmoid is usually used in multi-label classification tasks. For \texttt{GZFNO}, we also remove the post-processing that is only useful for single-label detection tasks. Except for the changes mentioned above, we retain all implementation details from the respective papers and run the models on the Guzheng\_Tech99 dataset.
	
	Table \ref{tab:compare} shows the frame-level precision, recall and F1-score of each method trained and tested on the Guzheng\_Tech99 dataset. Compared with the baseline methods, the proposed method achieves the highest scores in all metrics, confirming the effectiveness of our proposed model in the frame-level multi-label IPT detection task.
 
    \begin{figure}[t]
    \centering
    \includegraphics[width=8.3cm]{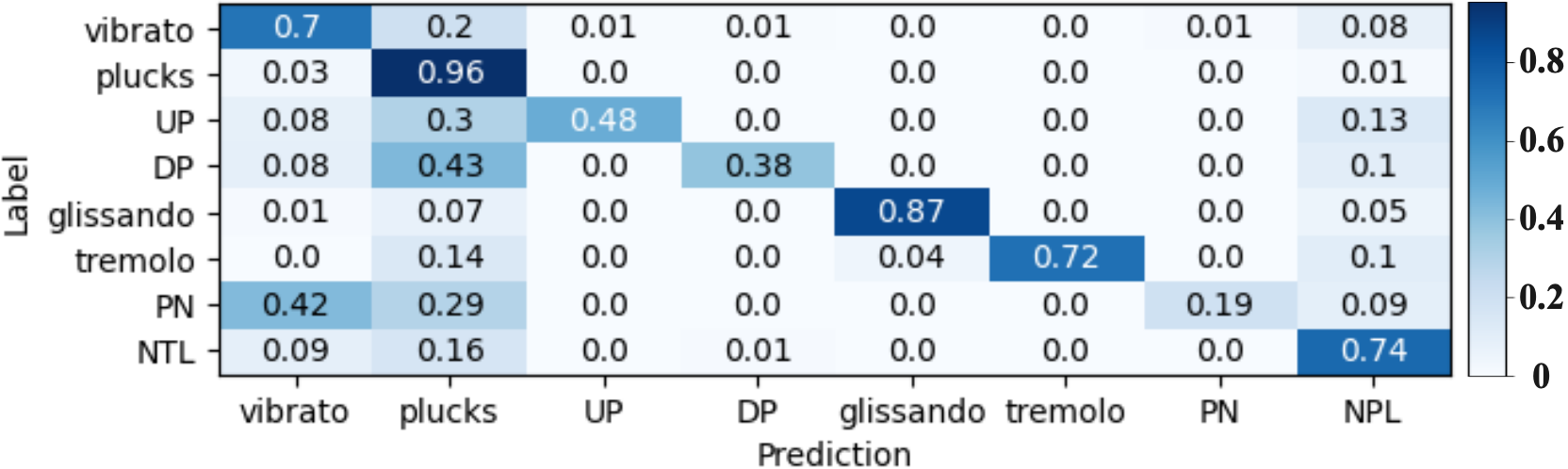}
    \caption{The confusion matrix of IPT detection using our model on Guzheng\_Tech99 dataset. ``NTL" and ``NPL" denote ``No True Label" and ``No Predicted Label", read \cite{heydarian2022mlcm} for details.}
    \label{fig:CM}
    \end{figure}
    
	\subsection{Comparing between Each IPT}
    To analyze individual IPT types in depth, we create a 2D confusion matrix using {\itshape mlcm} \cite{heydarian2022mlcm}. 
    The number on the main diagonal indicates the proportion of true positives for the label class.
    A high value in a row indicates that the particular pair of IPTs was frequently confused by our model. 
    As shown in Fig.\ref{fig:CM}, plucks has the highest true positive proportion, vibrato, {\itshape UP}, {\itshape DP}, and {\itshape PN} are often misclassified as plucks, and {\itshape PN} is particularly prone to misclassification as vibrato.
    Misclassifying {\itshape DP} as plucks also occurs frequently in other instruments \cite{wang2020playing, guqin}. It is mainly caused by the data imbalance, as plucks is more frequently used than other IPTs (see Table \ref{tab:statistic}). Besides, DP can be overlapped with plucks or mixed with tremolo (see Fig.\ref{fig:notation}), which can make accurate classification difficult.
    Furthermore, as shown in the spectrogram (Fig.\ref{fig:notation}), {\itshape PN} can be regarded as a special type of vibrato with only one pitch change, and there may be little differences in some players' performance between vibrato and {\itshape PN} when the note is extremely short. This similarity between PN and vibrato increases the likelihood of misclassifying PN as vibrato by our model.
    \section{Conclusion}
    In this paper, we formulate the IPT detection task as a frame-level multi-label classification problem. We create a new dataset, Guzheng\_Tech99, containing Guzheng recordings and onset, offset, pitch, IPT annotations of each note. A multi-scale network and a self-attention mechanism applied to the feature maps at the coarsest scale are designed to extract features from different scales and enhance the long-range feature extraction respectively. Our approach achieves 86.54\% in frame-level F1-score, outperforming the existing works by a large margin, which indicates its effectiveness in IPT detection. Although this work focuses only on Guzheng, the methodology can be applicable to other instruments. Future work will further verify its applicability on other instruments with rich IPTs, such as violin \cite{yang2017computational} and Chinese bamboo flute \cite{wang2022adaptive}. We will also expand the model to the note-level IPT detection further.

	
	\bibliographystyle{IEEEbib}
	\bibliography{strings,refs}
\end{CJK*}
\end{document}